\documentclass[]{spie}  

\usepackage{amsmath,amsfonts,amssymb}
\usepackage[utf8]{inputenc}
\usepackage{graphicx}
\usepackage[colorlinks=true, allcolors=blue]{hyperref}
\usepackage{cleveref}
\usepackage{subcaption}
\usepackage{bookmark}
\usepackage{blindtext}
\title{Cryogenic characterisation for the Nulling Interferometry Cryogenic Experiment (NICE)}

\author[a]{Jonah T. Hansen}
\author[a]{Emilie Bouzerand}
\author[a]{Adrian M. Glauser}
\author[b]{Bastien Rouz{\'e}}
\author[a]{Noah Stocker}
\author[c]{Walter Bachmann}
\author[c]{Marcel Baer}
\author[b]{Cindy Bellanger}
\author[a]{Thomas Birbacher}
\author[a]{Germain Garreau}
\author[a]{Julio Pino-Jiménez}
\author[b]{Jerome Primot}
\author[a]{Eckhart Spalding}
\author[a,d]{Sascha P. Quanz}
\affil[a]{ETH Zurich, Institute for Particle Physics and Astrophysics, Wolfgang-Pauli-Str. 27, 8093 Zurich, Switzerland}
\affil[b]{ONERA–The French Aerospace Lab, F-91761 Palaiseau, France }
\affil[c]{ETH Zurich, Department of Physics, Auguste-Piccard-Hof 1, 8093 Zurich, Switzerland}
\affil[d]{ETH Zurich, Department of Earth Sciences, Sonneggstrasse 5, 8092 Zurich, Switzerland}

\authorinfo{Send correspondence to J.T.H, E-mail: johansen@phys.ethz.ch}

\begin{document} 
\maketitle

\begin{abstract}
The Nulling Interferometry Cryogenic Experiment (NICE) is an experimental testbed for the beam combiner of the Large Interferometer For Exoplanets (LIFE) space mission. Until now, progress on NICE has been confined to an ambient bench, where we have recorded progress in deep ($<10^{-5}$) nulls at wavelengths between 4 and 5\,µm at 300\,K. However, the ultimate goal and requirement of NICE is to repeat these measurements at the sensitivity levels expected for a planetary system, requiring deep cryogenic conditions at 15\,K.
 
Here, we describe the ``Ice Cube''  cryostat, a small version of the future NICE cryostat that is used for component and subsystem level cryogenic testing. This is interfaced with a measurement setup using a segmented aperture interferometer and a wavefront sensor. We will also describe the testing campaign for understanding the material and mounting challenges that will be faced when translating the warm bench to cryogenic operations.

\end{abstract}


\section{INTRODUCTION}
\label{sec:intro}  

The LIFE space mission \cite{Quanz-2022-ID17,Glauser-2024-ID1} aims to be a transformative space observatory that can find and characterise Earth-like exoplanets in the habitable zones of nearby stars. To achieve this goal, it is envisioned as a mid-infrared (MIR) space-based nulling interferometer, working between 4 to 18.5\,µm. Such a mission, based on the heritage of the \textit{Darwin}\cite{Kaltenegger-2005-ID61,Cockell-2009-ID56} and TPF-I\cite{Lawson-2007-ID57} mission concepts, still requires a large amount of technology development -- particularly in the vein of cryogenic optics, mounts and actuators for an ultra-stable space mission. While we build on the experience gained from instruments such as JWST/MIRI \cite{Rieke-2015-MIRI}, which work at similar wavelength and temperature regimes, the necessary stability of phase and amplitude are more stringent\cite{Birbacher-Hansen-2026} and hence require a dedicated testbed.

This is one of the primary purposes of NICE - the Nulling Interferometry Cryogenic Experiment \cite{Ranganathan-2024-ID9,Birbacher-Hansen-2026}. NICE is a laboratory testbed designed to replicate the LIFE nulling instrument, covering the same wavelength range (4-18.5\,µm), sensitivity (ability to detect the equivalent of an Earth-twin at 10\,pc) and contrast (1$\times10^{-5}$). To achieve these parameters, specifically the sensitivity at longer wavelengths, NICE (and by proxy the LIFE instrument) must be cryogenic at temperatures between 15 and 50\,K; these values stem from approximations of the amount of thermal emission expected from the telescope and instrument.

To date, most progress on NICE has revolved around an ambient precursor bench to define and verify the optical layout, identify key challenges and iterate on opto-mechanical requirements. Updates on the progress of this bench, including measurements of a deep $7\times 10^{-6}$ null at 4.7\,µm, can be found in the recently published peer-review study \cite{Birbacher-Hansen-2026} and the dedicated update at this SPIE meeting (Hansen et al., proc. 14148-26 \cite{Hansen_NICE_SPIE_2026}). However, with an ambitious aim of launching LIFE by 2040 (see Glauser et al., proc. 14148-64 \cite{Glauser_SPIE_2026}), urgent progress is needed to move NICE towards a cryogenic environment.

To assist with this, NICE requires the ability to rapidly characterise infrared components, including optics, mounts and small sub-assemblies, without requiring weeks between cool-down. As such, we have developed the ``Ice Cube'' cryostat - a small, modular unit that can provide a vacuum and $<$\,15\,K environment for optical testing with a relatively quick turnaround. In this proceeding, we will describe the design and requirements of the cryostat and some of its initial performance. We will also detail the initial optical measurement system for characterising minute optical deformations during cooldown, and some of the first tests of optical mounts. Finally, we end with an outlook on the required next steps for the cryogenic implementation of NICE.

\section{Ice Cube design and performance}
\label{sec:cryo_design}

\subsection{Design Drivers and Requirements}
\begin{figure}
    \centering
    \includegraphics[width=\linewidth]{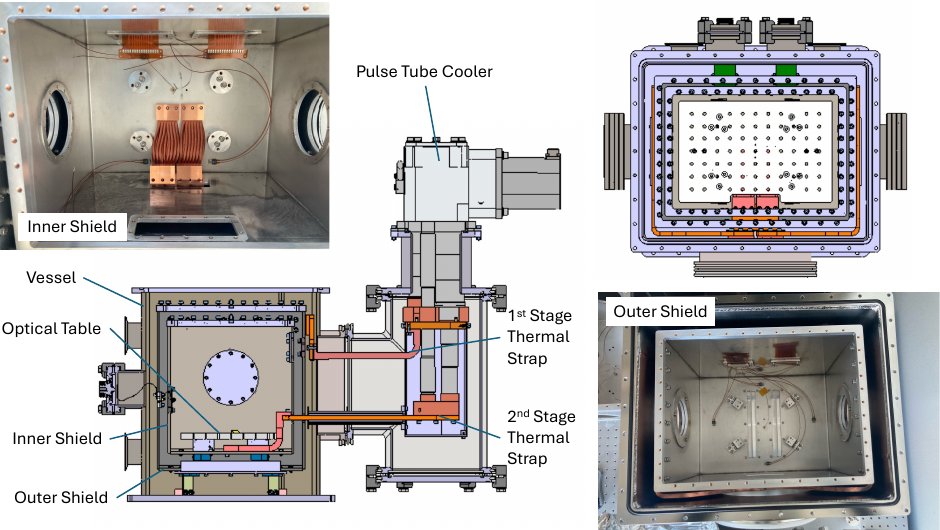}
    \caption{CAD schematic of the Ice Cube cryostat design, with photographs indicating the assembled components.}
    \label{fig:CAD}
\end{figure}

The ``Ice Cube'' cryostat was developed as a versatile cryogenic platform dedicated to the testing and characterisation of infrared optical components, with the possibility of accommodating small optical sub-assemblies. The main requirements are to provide a stable cryogenic environment for precision optical testing, to ensure direct access to the cold optical volume, and to allow integration of future test items whose detailed configuration was not defined during the design phase. A schematic of the cryostat is shown in \cref{fig:CAD}.

A core philosophy of the design is the ability to rapidly cycle, open, and close the system without dismantling the primary cooling loop. To achieve this, the Pulse Tube Cooler (PTC) is mounted entirely on one side of the vacuum vessel rather than on top. While this lateral placement slightly reduces absolute cooling efficiency due to longer thermal paths, it completely uncouples the cooler from the main access ports. All vacuum pump-out ports and electrical feedthroughs are segregated to the side opposite the PTC, leaving the remaining three covers of the cube entirely unobstructed. Users can directly access the cold optics and the optical table by simply removing these three covers without disturbing the cryogenic cooling system.

To ensure maximum adaptability for unknown test configurations, the internal aluminium optical table features a regular grid bolt pattern. For optical testing, two windows are integrated into the vessel, which include blank-off covers to completely block radiative parasitic loads when not in use. For initial tests, silica windows are implemented for visible light access and a reduction in thermal leakage. Future iterations may require the integration of MIR transparent glasses or fibre ports to allow for testing with infrared light; this is easily modified. A dedicated electrical feedthrough is also integrated to provide power and signal access for future test items.

\subsection{Cryogenic and Mechanical Design Concept}

 The cryogenic architecture is based on a two-shield configuration enclosed inside a stainless-steel vacuum vessel. The vessel is electropolished to reduce radiative heat transfer from the ambient environment. The outer thermal shield is cooled by the first stage of the PTC and operates at approximately 100\,K. Its role is to intercept the radiative heat load from the vessel before it reaches the colder components such that the thermal gradient across the inner shield is reduced. The second thermal shield is cooled by the second stage of the PTC and operates at approximately 10\,K. The thermal connection is strategically placed at the bottom of the shield so that the localised cold spot at the strap interface is completely hidden from the field of view of the tested optics.

The position of the PTC requires a dedicated thermal transfer solution. High-conductivity copper straps are used to transport cooling power from the cooler to the shields while minimising thermal losses. The thermal connection to the outer shield is located at the top to satisfy accessibility constraints. For the inner shield, the thermal connection is strategically placed at its bottom so that the localised cold spot at the strap interface is completely hidden from the field of view of the tested optics.

The shields are mechanically supported using G10 blades, which provide both thermal isolation and flexibility during cool-down. The blade dimensions are adapted to the different interfaces: those between the vessel and the outer shield are longer than those between the outer and inner shields due to the larger differential deformation due to cryogenic temperature that must be accommodated.

The orientation of the G10 blades follows the natural contraction directions of the cryogenic structures. By positioning the blades perpendicular to the shield diagonals, the cool-down deformation occurs with minimal mechanical constraint. This results in a nearly stress-free contraction of the shields and optical table, producing a highly uniform deformation pattern and maintaining approximately zero in-plane displacement at the centre of the optical table.

The optical table is manufactured from aluminium and is thermally connected to the inner shield through dedicated aluminium interfaces. It is intentionally not directly connected to the PTC to avoid creating a local cold region. Instead, cooling is distributed through four thermal connection points, producing a more uniform thermal environment across the optical surface.

\subsection{Cryostat performance}

\begin{figure}
    \centering
    \includegraphics[width=0.8\linewidth]{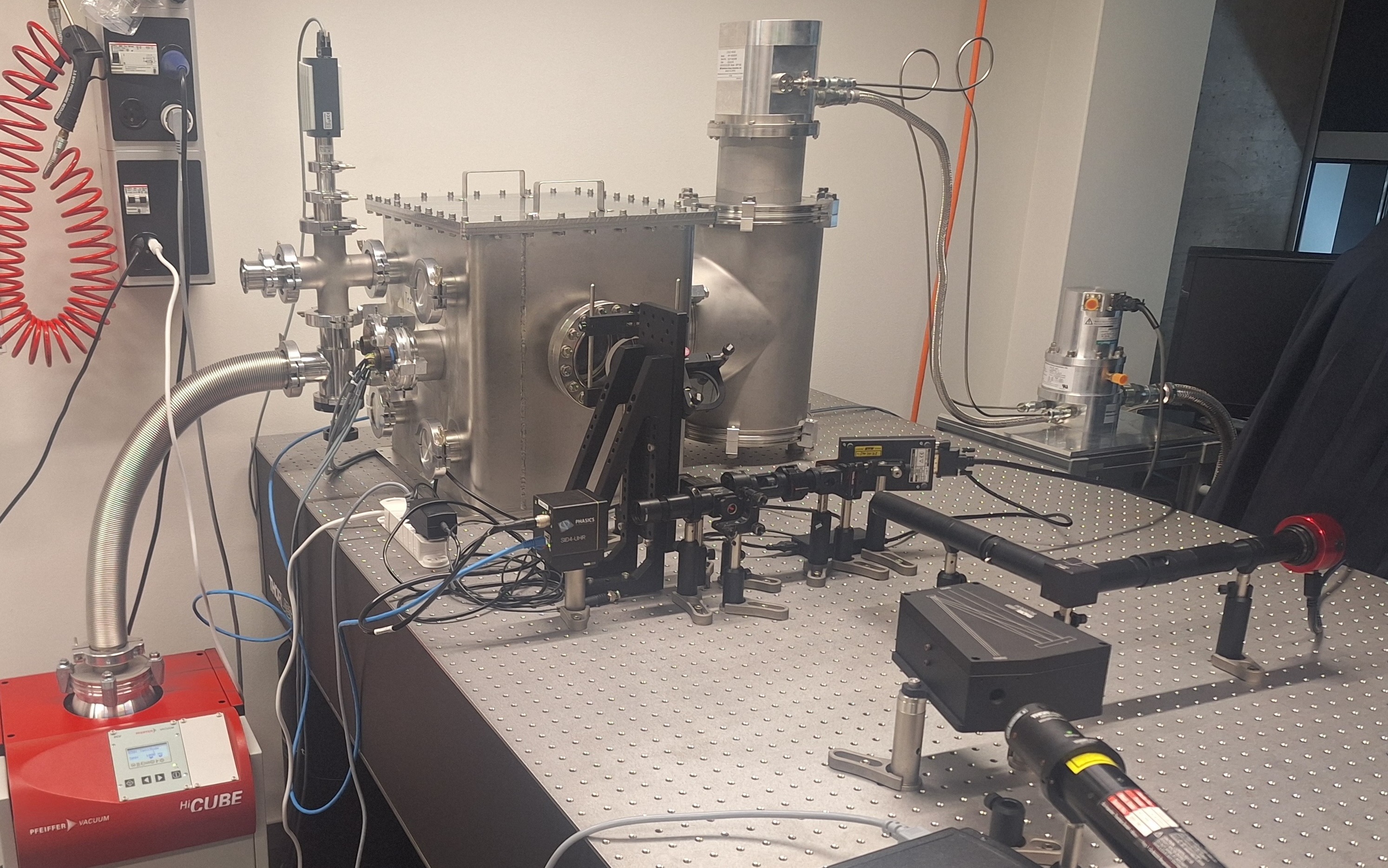}
    \caption{The assembled Ice Cube cryostat, along with the optical characterisation setup described in \cref{sec:measurement} and \cref{sec:tests}}
    \label{fig:Ice_cube_assembled}
\end{figure}

The fully assembled Ice Cube is shown in \cref{fig:Ice_cube_assembled}, along with the optical measurement setup described in \cref{sec:measurement} and \cref{sec:tests}. 

The combination of the two-shield architecture and distributed thermal connections provides a highly uniform cryogenic environment. The inner shield and optical table operate near 10\,K with a gradient below 2\,K for the shield and below 0.05\,K at table, minimising thermally induced optical distortions.  Temperature stabilisation is achieved using temperature sensors and electrical heaters integrated into the cryogenic stages. This active control allows the operating temperature to be maintained according to the requirements of different experiments.

Initial cool-down indicates that the outer shield reached a temperature of 100\,K, inner shield 14.4\,K, and optical table 18.1\,K, along with a pressure of 1.5$\times 10^{-7}$\,mBar. However, while the inner temperatures are close to the targets, they are a few Kelvin warmer than predicted. Current investigations indicate that the PTC may not be providing the level of cooling expected, and that there may be unforeseen coupling from the heaters to the inner stage. Applying the fixes to both of these issues are ongoing, though they do not prevent the start of cryogenic testing in parallel. Detailed characterisation of the cool-down curve and control stability will follow.

\section{Optical characterisation principle}
\label{sec:measurement}

In conjunction with the Ice Cube cryostat, an optical measurement setup is needed to characterise the optics before, during and after cool-down. Such a system is not straightforward, in so much that small vibrations from the vacuum pump and PTC may dominate over the optical deformations. Furthermore, as NICE is an interferometer with a single-mode spatial filter, the key relevant optical aberrations are piston and tip/tilt modes; many wavefront sensors are not so sensitive to these low-order modes. 

\begin{figure}
    \centering
    \includegraphics[width=0.8\linewidth]{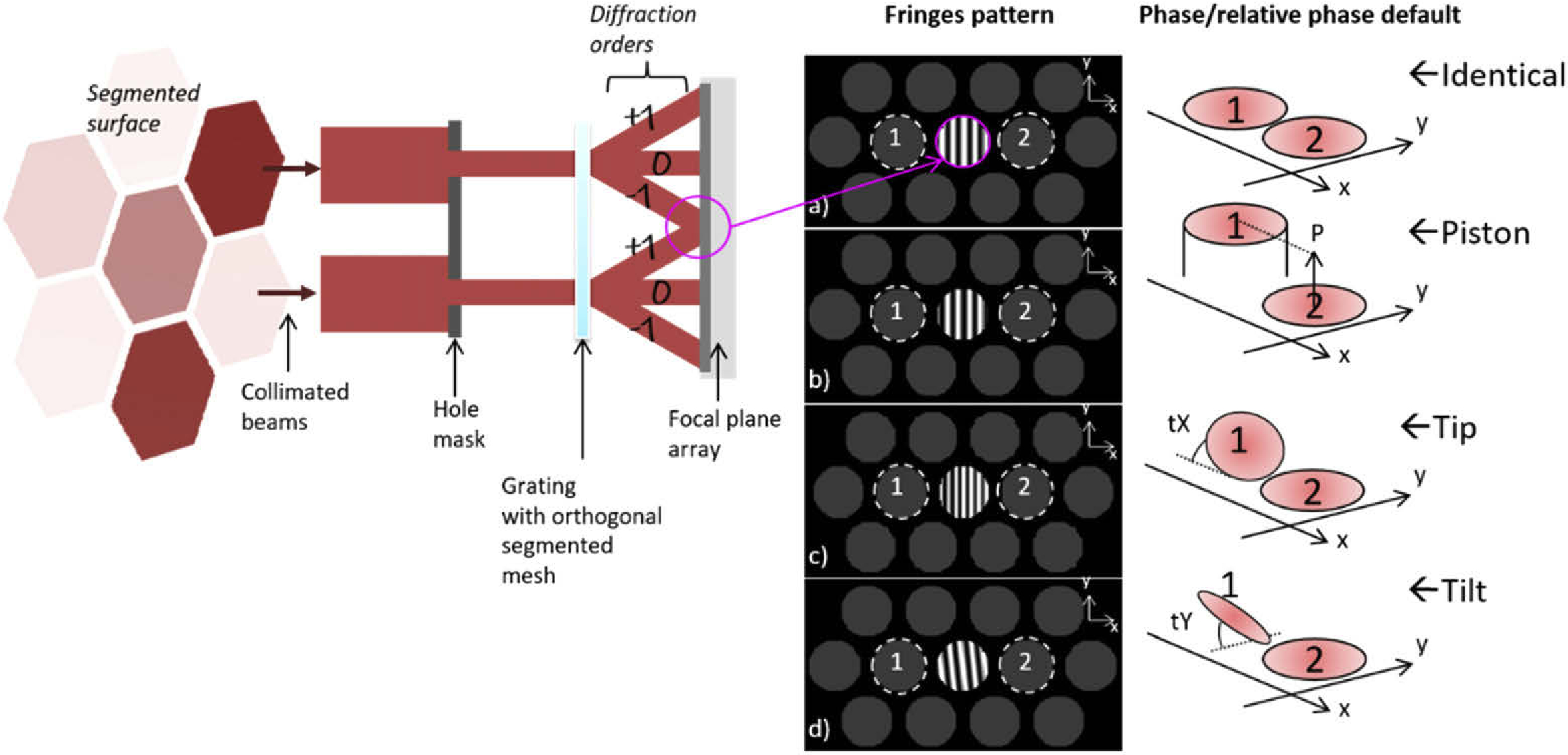}
        \caption{Principle of the PISTIL measurement, whereby piston, tip and tilt changes between sub-apertures are encoded in fringes. Taken from Rouz{\'e} et al. (2020)\cite{PISTIL2020}.}
        \label{fig:pistil_principle}
\end{figure}

\begin{figure}
        \centering
    \includegraphics[width=0.7\linewidth]{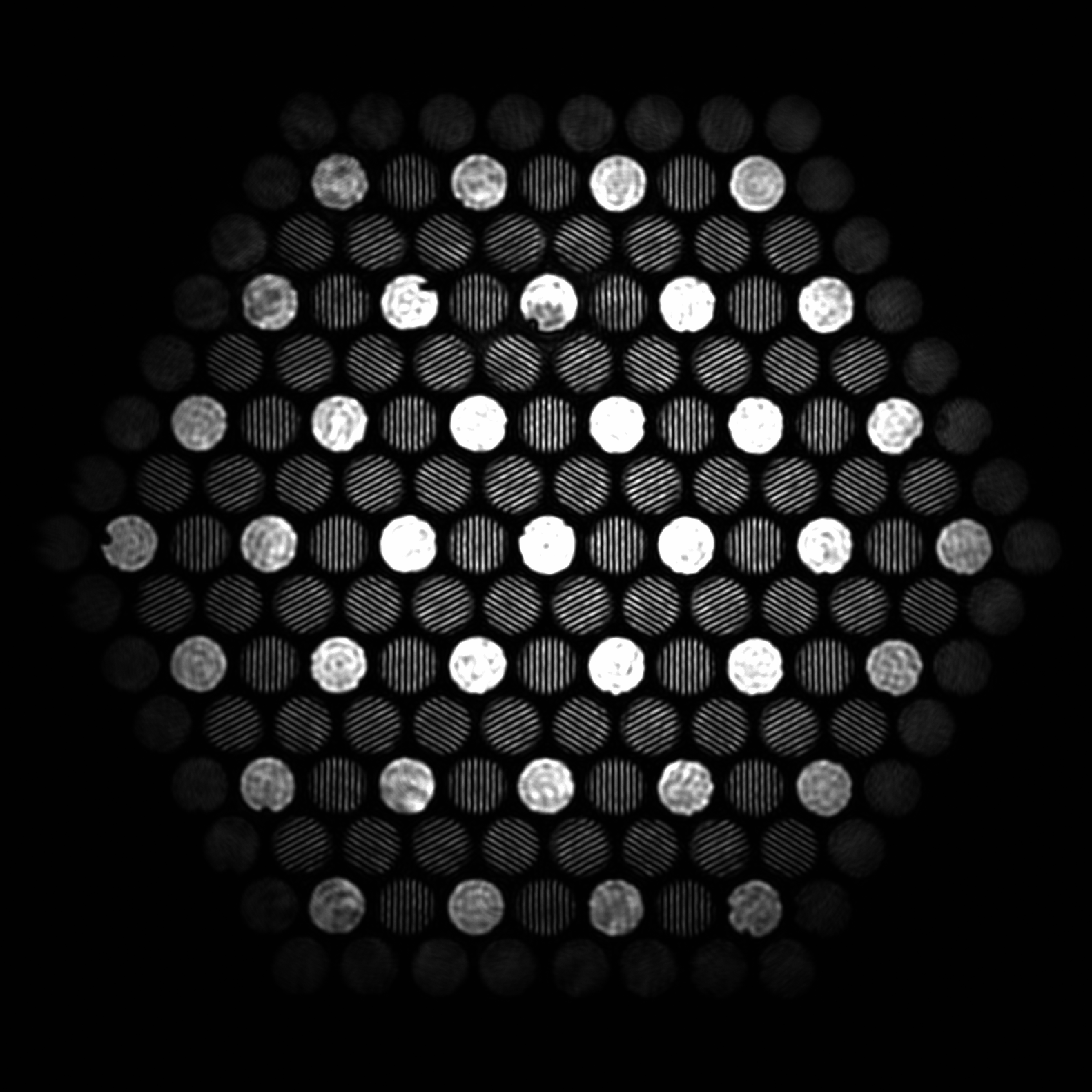}
    \caption{Example of a measured ``pistilogramme''}
    \label{fig:pistilogramme}
\end{figure}

Hence, we have adapted the PISTIL (PISton TIp and Tilt interferometry) measurement principle as detailed in e.g. Deprez et al. (2018)\cite{PISTIL2018}; Rouz{\'e} et al. (2020)\cite{PISTIL2020} and Rouz{\'e} (2021)\cite{rouze2021interferometrie}. First proposed to measure and co-phase hexagonal deformable mirrors from afar, the concept relies on analysing pair-wise fringes obtained by the first-order diffraction order of a segmented aperture. In this way, it works in a similar vein to both aperture-masking and lateral-shearing interferometric techniques. As the relevant piston and angle between two sub-apertures change, the fringes encode this information to be extracted. A schematic of the measurement principle and the relevant fringe encodings is found in \cref{fig:pistil_principle} and an example of a measured ``pistilogramme'' is shown in \cref{fig:pistilogramme}. 

To apply this technique to the deformations expected during cool-down for NICE, we include a mirrored annular reference mirror. The PISTIL setup then measures the deformation of an inner test mirror with respect to the outer reference mirror, which while not providing absolute measurements in piston, works to identify the relative movement between optic and mount.

\begin{figure}
    \centering
    \includegraphics[width=\linewidth]{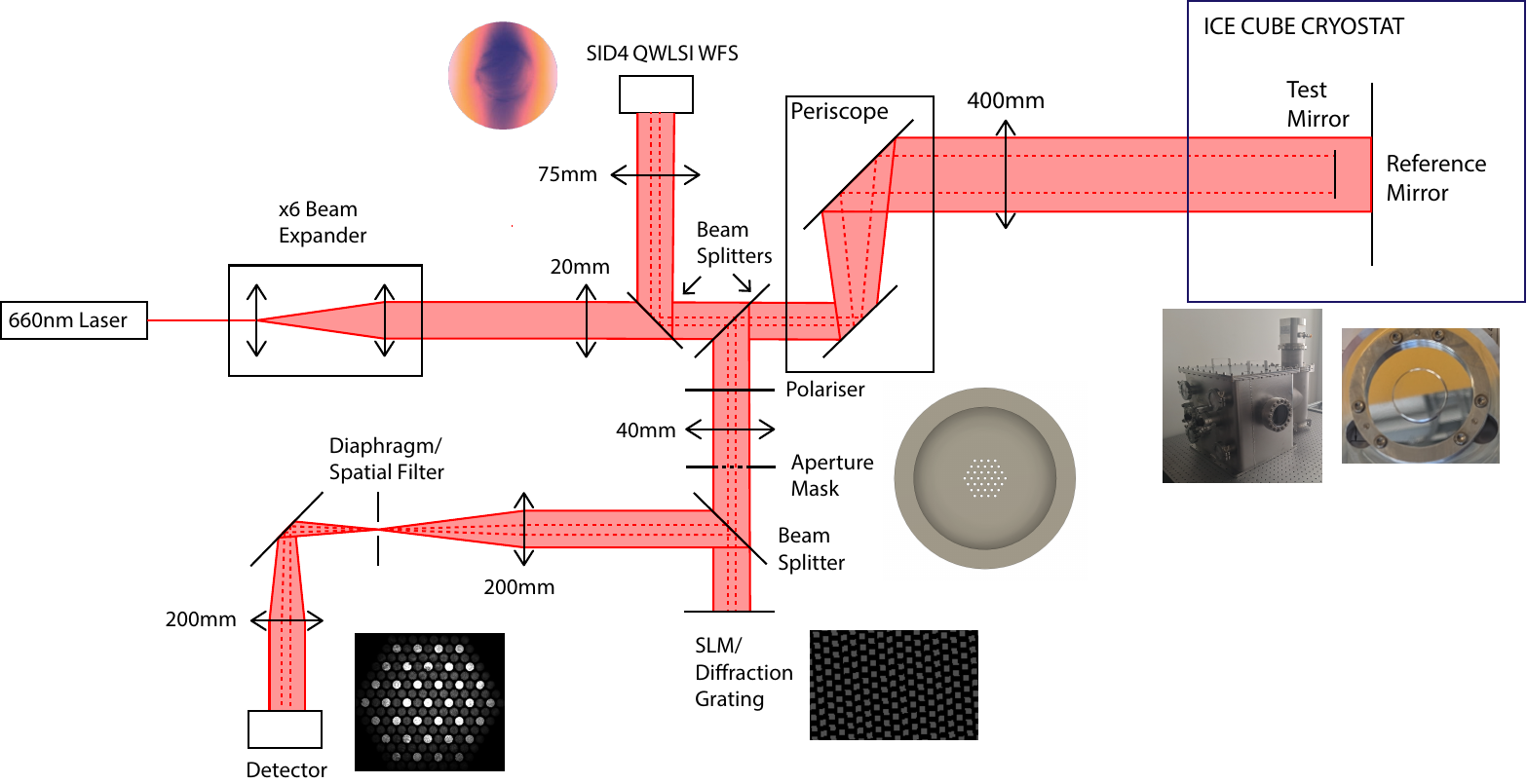}
    \caption{Optical diagram of the Ice Cube optical characterisation bench. An incoming 660\,nm laser incident on a reference and test mirror inside the cryostat is reflected back to two arms, one containing the PISTIL measurement, and the other with a SID4 QWLSI wavefront sensor from Phasics. More information is given in the text.}
    \label{fig:PISTIL_diagram}
\end{figure}

The full optical diagram is shown in \cref{fig:PISTIL_diagram}. A 660\,nm He-Ne laser is expanded via a series of beam expanders and directed into the Ice Cube cryostat window. The beam hits the annular reference and test optic simultaneously, before being reflected back along the same path. A beam splitter retrieves the outgoing beam to direct through a mask containing a hexagonal array of holes that define the sub-apertures being analysed. The beam is then incident on a two-dimensional diffraction grating to generate the interfering diffraction orders. In the original PISTIL setup, discrete gratings were manufactured; however, to allow modularity with the types of masks and aperture shapes we can analyse, we impose the diffraction grating through the use of a reflective spatial light modulator (SLM). Finally, we focus the beam to a diaphragm that removes the remaining high-order diffraction, before imaging onto a detector such that the first-order diffracted beams lie in-between the zeroth-order sub-aperture spots. All relevant planes, including the mirror and mask, are conjugated to avoid the effects of unwanted diffraction, and all measurements are taken with respect to a reference in order to mitigate system alignment errors. We invite the reader to look at the aforementioned references for more details on this measurement scheme. 

For our implementation, we utilise a 37-element hexagonal mask with hole radius 200\,\textmu m and separation 950\,\textmu m. This, combined with the magnification 0.1 beam expander, results in sampling wavefront segments 4\,mm in diameter. The sizing was chosen so that we can cleanly sample 1, 2 and 3~in mirrors. Furthermore, a grating period of 90\,\textmu m results in 9 fringes per sub-aperture, and a pixel size of 2.4\,\textmu m allows for a sampling of 9~pixels per fringe.

One potential issue with this technique is the limited capture range of the measurement, whereby movements greater than $\lambda/4$ (in double pass) will result in a phase wrapped degeneracy in piston, and tip/tilts larger than $G\lambda/p_g$ will result in the diffracted orders being blocked by the iris or filtered in post-processing. Here, $G$ represents the total magnification from the mirror to the detector (which in our current setup is 0.1), and $p_g$ is the grating period. One way to overcome the former issue is that of including a second, similar wavelength to perform two-wavelength interferometry \cite{PISTIL2020}, though at the cost of extra complexity and a second laser and camera. Another method is to perform data collection at a high enough cadence such that each individual measurement does not phase wrap, and then reset the reference every frame. This would result in less precision at the trade-off of an easier measurement.

While the PISTIL technique can provide estimates of the low-order modes, extracting information on higher-order aberrations such as a change in focus, coma or astigmatism is more challenging. To cover these higher-order aberrations, we have simultaneously installed a SID4 quadri-wave lateral shearing interferometer (QWLSI) wavefront sensor from Phasics into the beam train, which is picked off in the return reflection via another beam-splitter. This sensor is designed to see how the test mirror changes in these higher-order terms from before and after cool-down during the PISTIL measurements. 

We note here that the measurement has so far assumed reflective optics. In principle, refractive optics are also characterisable through a double pass: transmitting the beam through the full cryostat (including the back-side window), before returning back through the cryostat through the use of an external mirror.

To characterise an optic, therefore, we take a time series measurement of the optics both before, during and after cool-down, and measure the relative deformations between reference and test optic as a function of temperature. Based on these measurements, we can then identify whether certain mounting techniques result in too much differential deformation and verify STOP (structural-thermal-optical-performance) analysis simulations. We also anticipate extending these measurements beyond single components to subsystems such as the beam combiner or achromatic phase shifter, whereby we can measure the thermo-mechanical deformation as it compares to the optical requirements.

\section{Preliminary optical tests}
\label{sec:tests}

\subsection{Measurement verification}
\label{sec:verification}

\begin{figure}
    \centering
    \includegraphics[width=0.9\linewidth]{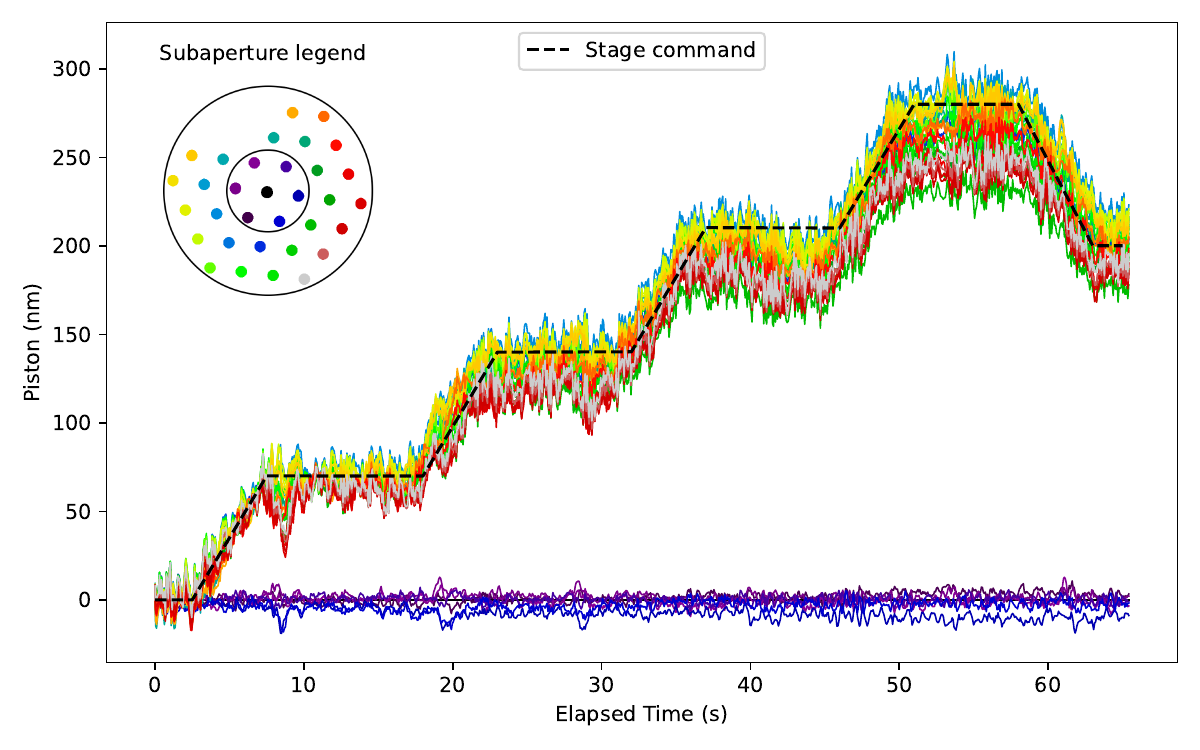}
    \caption{Piston of various sub-apertures across a pupil spanning two mirrors as a function of time. During the measurement, a linear stage moved one mirror with respect to the other; this is shown in the dashed black line. A legend for the colours representing each sub-aperture, and displaying which ones belong to which mirror, is shown in the upper left.}
    \label{fig:verification_piston_plot}
\end{figure}

To ensure the measurement performs, we conducted a verification test with two standard polished mirrors, one mounted on a post located on a one-axis translation stage. The stage was then set to move, relatively slowly, a set number of nanometres during the measurement, with the aim of recovering the piston movement. The reference was updated every frame, and the cumulative recovered piston for each sub-aperture is shown in \cref{fig:verification_piston_plot}. Over-plotted in dashed lines is the stage's reported position, though the exact timestamps have been estimated. Note that the absence of the sub-apertures in the upper-left corner is due to the positioning of the mounting post.

What we can see is that the measurement did a very good job at capturing the relative motion of the mirrors, albeit with some noise and drift between sub-apertures. A portion of the drift could be accounted for by the relatively long mounting post, which may have resulted in some tip/tilt changes in the mirror as it moved. This is corroborated by the fact that the sub-apertures closest to the pole have the least amount of deviation from the stage position. Ultimately, we see that the measurement is relatively accurate, with the greatest deviation being about 20\% from the true position, and has a precision of up to 7\,nm RMS in a lab environment with rather high seeing (being on a warm top floor of a nine story building).

\subsection{Initial optics during cool-down}

\begin{figure}
    \centering
    \includegraphics[width=0.7\linewidth]{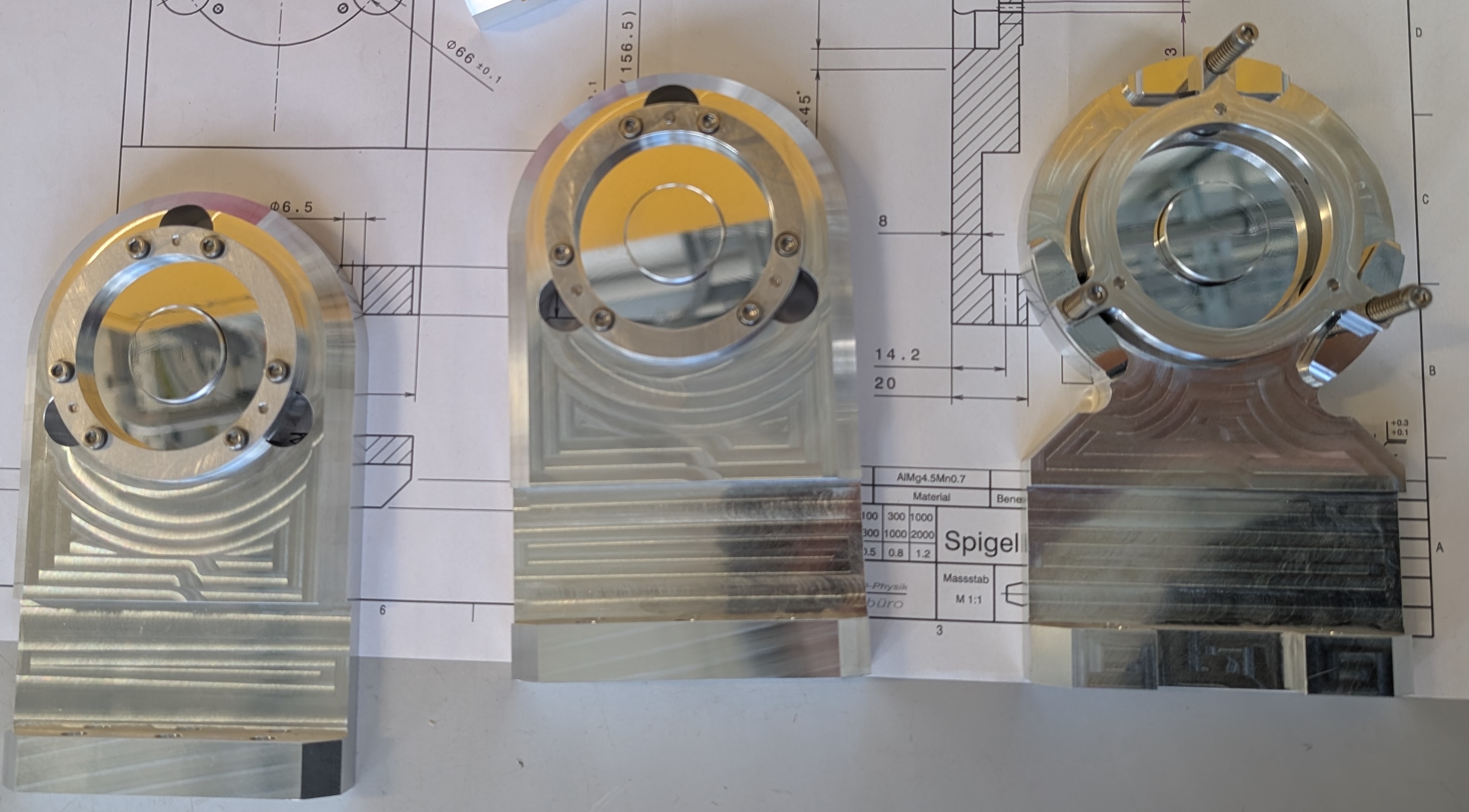}
    \caption{Three variations of an aluminium optic mounted to an aluminium structure, manufactured in-house at ETH Zürich. The first two involve directly mounting the optic to the structure, and the third utilises flexures for fine adjustments. All three utilise CNC milled mirror surfaces for the optical reflections.}
    \label{fig:workshop_mounts}
\end{figure}

For the first tests with the Ice Cube cryostat, we turn to a few options for mounting aluminium mirrors to aluminium mounts. The ETH Zürich Physics Workshop were able to manufacture a number of variations of a mounting system, with all involving one 25\,mm optic that is mounted from the back into the structure. The mount portion also contains a 50\,mm diameter mirrored annulus surface as is required by the PISTIL measurement. The optical surfaces were milled via a CNC machine with a diamond tool. Some of the components are shown in \cref{fig:workshop_mounts}. 

Of note are two variations: one is a direct mount, whereby the mirror is screwed directly into the mount after a CMM informed shim. The second utilises fine-threaded screws attached to flexures for minute, repeatable, µm-level adjustments in tip/tilt and position. Both need to be traded off in terms of their thermal performance, as while the latter is more promising when considering alignment alone, it may have greater thermal deformation.

We installed the directly mounted component into the cryostat, and set the PISTIL system measuring while the cryostat was cooled from 300 to 20\,K. Due to system constraints, we took data at a cadence of 10 frames every 30\,s over the course of the $>24$\,hr cooldown. We note here that the SID4 WFS has yet to be fully calibrated with respect to the setup, and so we do not present any results from that measurement arm here. 

\begin{figure}
    \centering
    \includegraphics[width=0.5\linewidth]{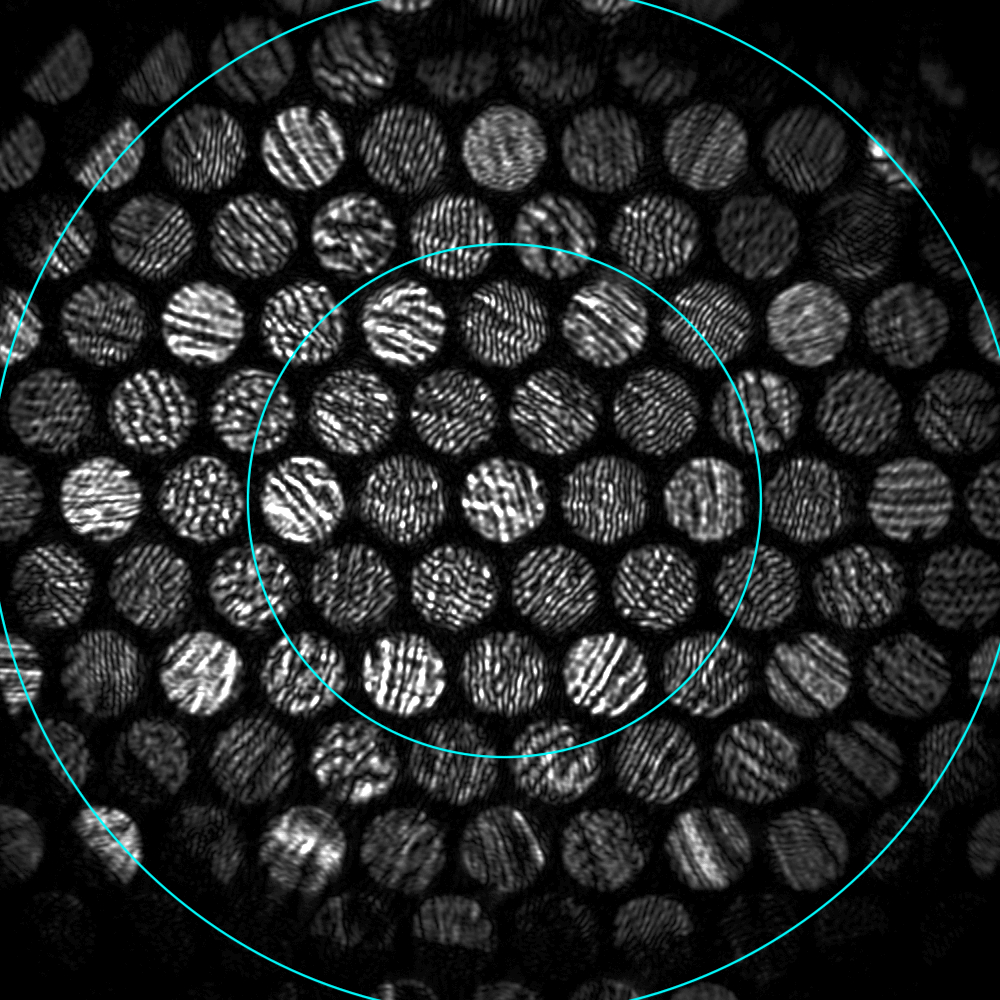}
    \caption{Pistilogramme as produced by the CNC-milled test mirrors. Overlaid are the mirror positions.}
    \label{fig:Al_pistilogramme}
\end{figure}

\begin{figure}
    \centering
    \includegraphics[width=0.9\linewidth]{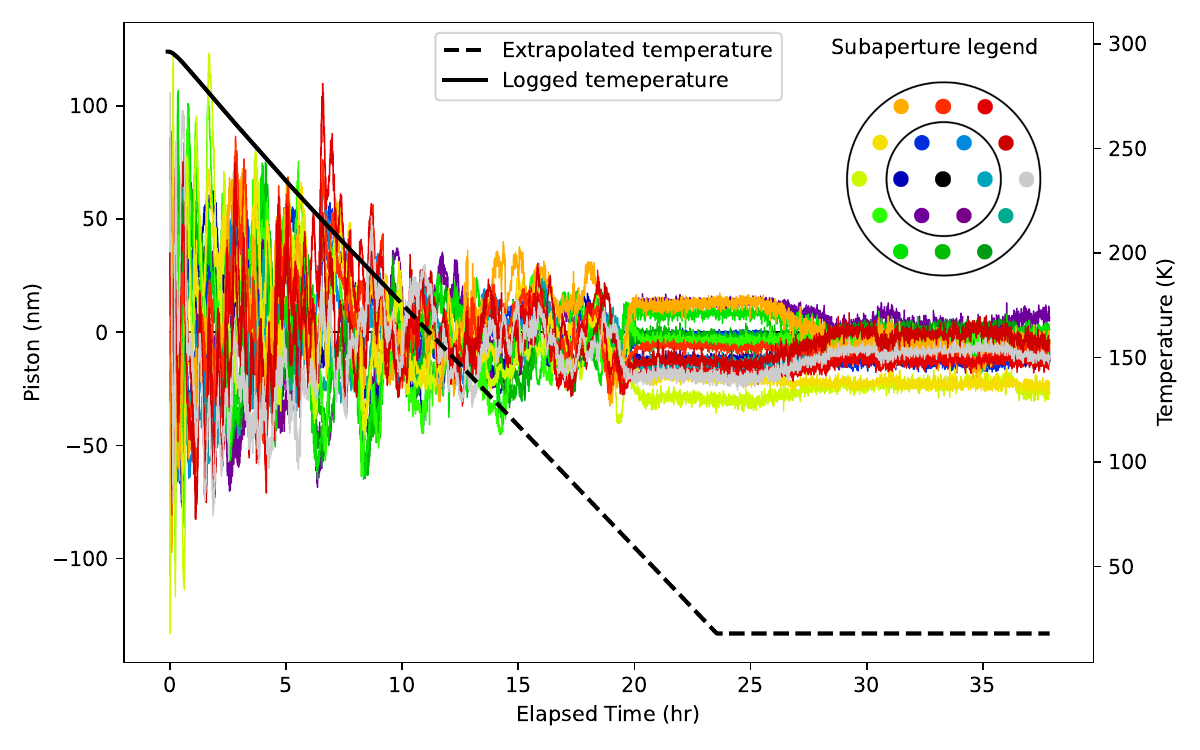}
    \caption{Piston of various sub-apertures across a pupil spanning two mirrors as a function of time, during which the mirrors were cooled from 300\,K to $\sim$20\,K. The temperature of the optical table is shown in a black line via the right y-axis; the temperature logging stopped half way through cool-down, and so the remaining data was linearly extrapolated to the final temperature. A legend for the colours representing each sub-aperture, and displaying which ones belong to which mirror, is shown in the upper left.}
    \label{fig:Ice_cube_piston_plot}
\end{figure}

A number of issues were found with the data collection. First, the mirrored surfaces were only marginally good enough for fringes as seen in the pistilogramme in \cref{fig:Al_pistilogramme}, which made it difficult to achieve the SNR for a wavefront measurement. Secondly, the capture ended up being too slow, with phase unwrapping becoming a problem in-between the 30\,s bursts. This can be seen in \cref{fig:Ice_cube_piston_plot}, where we plot the piston of the sub-apertures against the temperature of the cryostat using a single, non-updating reference. We can see that the piston fluctuates wildly during the cool-down, before settling down after the minimum temperature is reached. These fluctuations are caused by a combination of the deformation and an external noise source (such as seeing or vibrations), for which the relatively slow cadence was not able to disentangle. The relative referencing technique as described in \cref{sec:verification} did not work, again due to ambiguities with phase unwrapping between each 30\,s burst. The inconsistency in settling down with the temperature is due to a crash in temperature logging, whereby we then extrapolated the temperature curve to the final recorded temperature.

Therefore, future work will be spent on a number of improvements to the system:
\begin{enumerate}
    \item Increased optical quality of the test mirrors, to increase the SNR of the measurement.
    \item Real time data reduction to allow for a much higher data collection cadence, or a reduction in the spatial sampling of the pistilogramme.
    \item Characterisation of where the noise source comes from: whether it is seeing or vibrations.
    \item Stable temperature logging.
    \item Potential installation of a second wavelength to greatly increase the capture range.
    \item Alignment and calibration of the SID4 WFS.
\end{enumerate}
After improving the system, comparisons against the flexure mirror will occur in the near future.

\section{Future outlook}

The Ice Cube cryostat has nearly finished its commissioning run, with only a few small fixes (particularly surrounding the PTC) required before it reaches the theoretical specifications. Once fully functional, it will be used, along with the optical measurement scheme, to test the minute deformations induced through cryogenic cool-down of optical mounting systems, and how they relate to NICE's strict specifications of position and angle. The goal is to have a relatively quick turnaround between design, test and result to allow rapid iterations and quick convergence on a cryogenic opto-mechanical design.

After this, the cryostat will be used for subsystem level assemblies, along with use as a ``plug and play'' device to characterise the thermal performance of auxiliary technologies, such as photonic integrated circuits or cryogenic deformable mirrors. The accelerated characterisation and qualification of these mounts, technologies and subsystems will lead to insights into the design and specifications of the full NICE cryostat, along with strategies as to the implementation of the final instrument for the LIFE space mission.

\acknowledgments 
 
Part of this work has been carried out within the framework of the National Centre of Competence in Research PlanetS
supported by the Swiss National Science Foundation under grants 51NF40 182901 and 51NF40 205606. This work was also supported by the Swiss National Science Foundation (grant number 10004532) and by the Swiss State Secretariat for Education, Research and Innovation (SERI) / Swiss Space Office (SSO). This project was supported by Rudolf Bär via the ETH
Zurich Foundation. No AI tools were used in either the analysis or writing of this manuscript.

\bibliography{report} 
\bibliographystyle{spiebib2} 

\end{document}